\documentclass[manuscript]{aastex}

\shorttitle{{\em Kepler} False-Positive Elimination }
\shortauthors{Batalha et al.}

\begin{document}

\title{Pre-Spectroscopic False Positive Elimination of {\em Kepler} Planet Candidates}

\author{Natalie M. Batalha}
\affil{Department of Physics and Astronomy, San Jose State University,
        San Jose, CA 95192}
\email{Natalie.Batalha@sjsu.edu}

\author{Jason F. Rowe}
\affil{NASA Postdoctoral Program Fellow, NASA Ames Research Center,
        Moffett Field, CA 94035}

\author{Ronald L. Gilliland}
\affil{STScI, Baltimore, MD 21218}

\author{Jon J. Jenkins and Douglas Caldwell}
\affil{SETI Institute, Mountain View, CA 94043}

\author{William J. Borucki, David G. Koch, and Jack J. Lissauer}
\affil{NASA Ames Research Center, Moffett Field, CA 94035}

\author{Edward W. Dunham}
\affil{Lowell Observatory, Flagstaff, AZ 86001}

\author{Thomas N. Gautier}
\affil{Jet Propulsion Laboratory, Pasadena, CA 91109}

\author{Steve B. Howell}
\affil{NOAO, Tucson, AZ 85726}

\author{David W. Latham}
\affil{Harvard-Smithsonian, CfA, Cambridge, MA 02138}

\author{Geoff W. Marcy}
\affil{University of California, Berkeley, CA 94720}

\and

\author{Andrej Prsa}
\affil{Villanova University, Villanova, PA 19085}

\begin{abstract}

Ten days of commissioning data (Quarter 0) and thirty-three days of science data (Quarter 1) yield instrumental flux timeseries of $\sim$150,000 stars that were combed for transit events, termed {\em Threshold Crossing Events} (TCE), each having a total detection statistic above 7.1-$\sigma$. TCE light curves are modeled as star+planet systems.  Those returning a companion radius smaller than $2R_J$ are assigned a KOI ({\em Kepler Object of Interest}) number.  The raw flux, pixel flux, and flux-weighted centroids of every KOI are scrutinized to assess the likelihood of being an astrophysical false-positive versus the likelihood of a being a planetary companion.  This vetting using {\em Kepler} data is referred to as data validation.  Herein, we describe the data validation metrics and graphics used to identify viable planet candidates amongst the KOIs.  Light curve modeling tests for a) the difference in depth of the odd- versus even-numbered transits, b) evidence of ellipsoidal variations, and c) evidence of a secondary eclipse event at phase=0.5.  Flux-weighted centroids are used to test for signals correlated with transit events with a magnitude and direction indicative of a background eclipsing binary.  Centroid timeseries are complimented by analysis of images taken in-transit versus out-of-transit, the difference often revealing the pixel contributing the most to the flux change during transit.  Examples are shown to illustrate each test. Candidates passing data validation are submitted to ground-based observers for further false-positive elimination or confirmation/characterization.

\end{abstract}


\section{Introduction}
\label{sec:intro}
The {\em Kepler} science team is reporting on its first planet discoveries (\citealp{koi7,borucki10,koi17,koi18,koi97}) in this volume -- discoveries discerned from data taken in the first weeks of science operations. The progression from pixels to planet detection passes through numerous stages, from target selection, data collection, and aperture photometry to transiting planet search, data validation, and follow-up observations.  In this progression, data validation (DV) and follow-up operations (FOP) are both tasks concerned, in part, with the vetting of false positives.  They are distinct, however, in that DV performs false-positive elimination from diagnostics that can be pulled out of the {\em Kepler} data itself, while the FOP relies on additional observations such as moderate and high-precision spectroscopic radial velocities \citep{fop}.  Automated DV is under development in the Science Operations Center at NASA Ames Research Center and will soon provide pipeline generation of metrics, reports, and graphics that the science team will use to sift through the thousands of transit events that {\em Kepler} expects to detect.  

These, our first discoveries, were scrutinized in non-pipeline fashion with some metrics taken from the pipeline DV, and others developed in real time to get the job done as efficiently as possible to support the 2009 ground-based observing season -- the first since launch.  Many of the tools that sprang up out of this effort have fed back into the DV development, ensuring it will be a powerful analysis pipeline for future processing.  A full description of the DV pipeline currently under development can be found in \citet{pipeline}.  

This communication describes the pre-pipeline DV tools applied to this first analysis of science data.  An event flagged by the {\em Transiting Planet Search} (TPS) pipeline as having transit-like features with a total detection statistic greater than 7.1-$\sigma$ is termed a {\em Threshold Crossing Event} (TCE).  Each TCE light curve is modeled, and those returning a companion radius $< 2 R_J$ are assigned a {\em Kepler Object of Interest} (KOI) number.  Only those that pass the DV tests described herein are submitted to the follow-up observers for spectroscopic vetting, confirmation, and characterization.  The objective is to eliminate as many of the false-positives as possible to efficiently utilize the limited amount of telescope time available for follow-up observations.  We note that elimination of a TCE (i.e. transit detection) as a viable planet candidate does not imply that the target itself is no longer observed or no longer subjected to transit searches.   Moreover, a light curve may yield multiple threshold-crossing events (TCEs).  Each is considered independently with regards to the planet interpretation.

The large majority of false-positives are caused by either grazing or diluted eclipsing binaries (\citep{brown03}), the latter being the more likely.  Dilution occurs when light from a nearby star falls within the photometric aperture of the foreground star, where ``nearby'' can be either a true physical companion to the star or a chance projection on the sky.  We refer to the latter scenario as a Background Eclipsing Binary (BGEB).  The photometric and astrometric precision of the {\em Kepler} photometer affords us unprecedented opportunities to vet out the false positives from the flux and photocenter timeseries themselves.   

Section~\ref{sec:obs} gives a brief summary of the data, from acquisition to transit detection.  Section~\ref{sec:modeling} describes the metrics used to identify grazing and diluted EBs derived from light curve modeling, and Section~\ref{sec:centroid} addresses the analysis of the photocenter variations.  The stars used to exemplify the various techniques will be referred to by their KOI designation as well as the {\em KeplerID} archived in the {\em Kepler Input Catalog}\footnote{\url{http://archive.stsci.edu/kepler}} (KIC).

\section{Observations, Light Curves, and Transit Detection}
\label{sec:obs}
The analysis is based on two sets of data: 1) a 9.7-day run, May 2 through May 12, 2009, during the commission period to measure the initial photometric performance of the instrument, and 2) the first 33.5 days of science operations, May 13 through June 17, 2009, the end of which is marked by a $90^\circ$ quarterly roll of the spacecraft about the optical axis \citep{haas}.  The former data set is referred to as Quarter 0 (Q0), while the latter is referred to as Quarter 1 (Q1).  All relatively uncrowded stars brighter than $Kp=13.6$ (where $Kp$ is the apparent magnitude in the {\em Kepler} passband) make up a list of 52,496 targets observed during Q0.  The Q1 target list contains 156,097 stars, $93\%$ of which were selected based on metrics that address the detectability of terrestrial-size planets \citep{targetselection} -- metrics dependent on stellar properties (e.g. effective temperature, surface gravity, and apparent magnitude).  The intersection of the two lists contains 45,742 targets. The aperture photometry, data conditioning, and transit search algorithms are described in \citet{pipeline}. A discussion of the resulting flux timeseries is given in \citet{longcadence}.

The transiting planet search applied to all stars yielded several thousand TCEs, some of which were triggered by artifacts in the light curves and single strong outliers.  This was mitigated by comparing the $\chi^2$ statistic of a transit model fit to the folded light curve against the $\chi^2$ statistic of a linear fit to the same.  If the two are statistically indistinguishable, the TCE is disregarded.  A constraint on the transit duration (1-hour) is used to isolate TCEs triggered by strong outliers in the flux timeseries. 

The transit model provides an estimate of the companion radius.  TCEs with a companion radius larger than $2R_J$ are also disregarded.  Such large radii are likely associated with late M-dwarfs and are not considered high priority planetary candidates.  This leaves approximately $10^3$ TCEs.

\section{Light Curve Modeling}
\label{sec:modeling}
We use the analytic expressions of \citet{man02} to model the transit of the planet. For our initial fits we use \citet{cla00} V-band nonlinear limb darkening parameters for the star as an approximation for the {\em Kepler} bandpass. Values for the stellar radius ($R_{\star}$), effective temperature ($T_{\rm eff}$) and surface gravity ($\log g$) are retrieved from the KIC. The stellar mass ($M_{\star}$) is computed from $R_{\star}$ and $\log g$. With $M_{\star}$ and $R_{\star}$ fixed to their initial values a transit fit is computed to determine the orbital period,
phase, orbital inclination, and planetary radius, $R_p$.  The best fit is found using a Levenberg-Marquardt minimization algorithm \citep{pre92}.

In the case of an eclipsing binary where the radii and temperature of the two components are slightly different, the depths of the eclipses imprinted on the light curve will be different. Every other transit will have a different depth. When at least 2 transits are present in the light curve, the transit model is recomputed for the odd and even numbered transits where only the companion radius is allowed to vary. If the companion radii differ by more than 3-$\sigma$ then the TCE is rejected.  

\begin{figure}
\begin{center}
\includegraphics{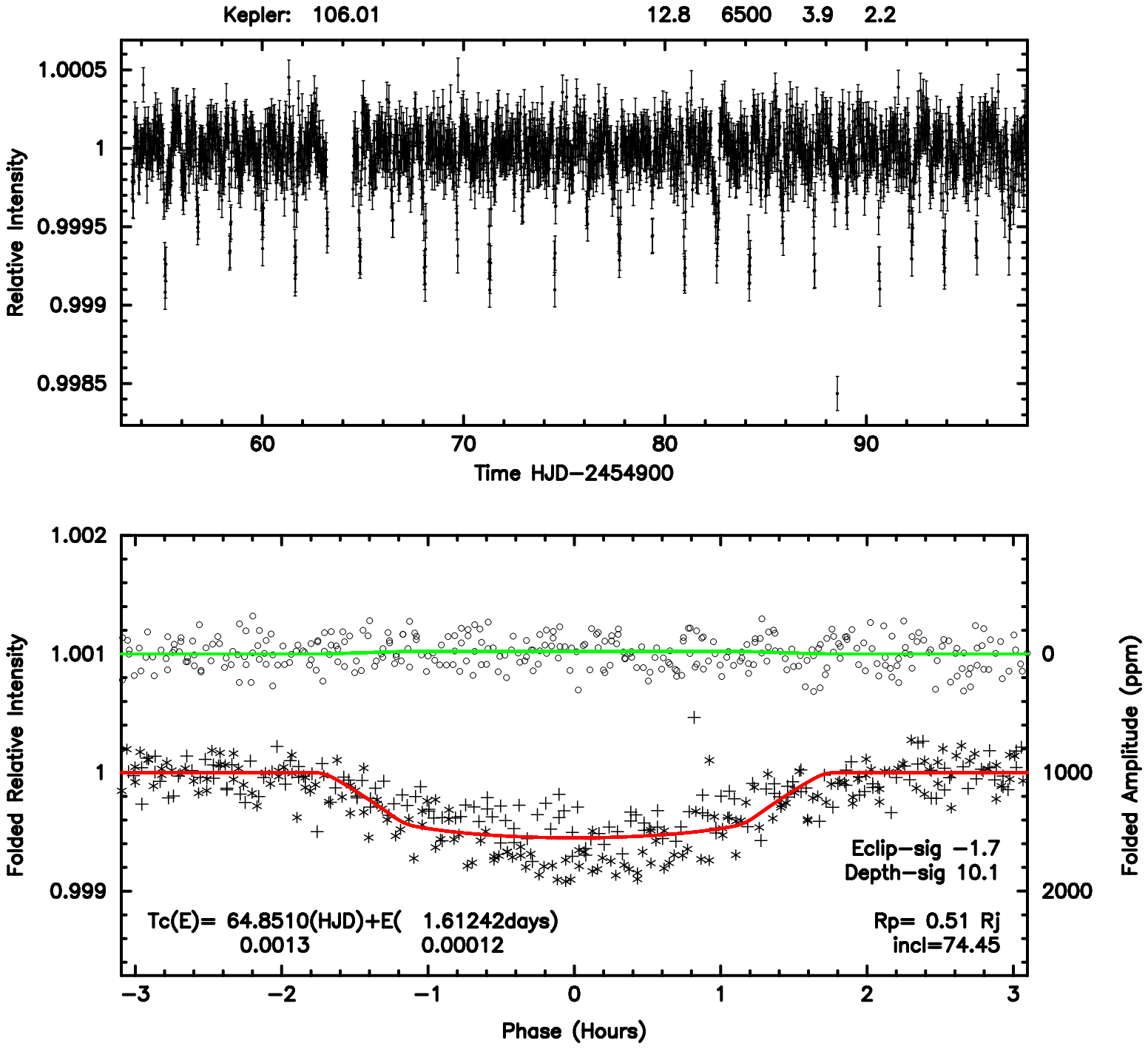}
\end{center}
\caption{The detrended light curve of KOI-106 (upper panel) is plotted against HJD (days).  The phased light curve is shown (lower panel) together with a model (solid line) that assumes the transit features are due to a planetary companion (resulting radius, $R_p=0.48 R_J$).  The region of the light curve centered on phase=0.5 is shown in the bottom panel (circles) and plotted in units of ppm (parts per million) as indicated on the axis on the right.  Stellar parameters ($Kp$, $T_{eff}$, $\log(g)$, and $R_*$ are listed at the top of the figure. The metrics $R_p$, Eclip-sig, and Depth-sig are defined in Section~\ref{sec:modeling}.  Modeling the odd (+) and even (*) numbered transits independently yields a significant difference in the companion radii (10-$\sigma$ difference) suggesting that this system is likely a diluted EB at twice the orbital period.}
\label{fig:oddEven}
\end{figure}

Figure~\ref{fig:oddEven} illustrates such an example. The upper panel shows the unfolded Q0+Q1 light curve of KOI-106 (KeplerID 10489525) -- a somewhat evolved F-type star according to the KIC.  Transit modeling provides an estimate of $0.48 R_J$ for the companion radius, $1.61236 \pm 0.00013$ days for the orbital period, and 74.06$^\circ$ for the inclination (shown in the lower panel).  The difference between the modeled companion radii utilizing first the odd transits and then the even transits, is given as depth-sig$=10.1\sigma$.  The folded light curve is shown in the bottom panel where the odd transits are plotted with asterisks and the even transits are plotted with plus symbols.  The difference in the depths of the odd versus even transits is clear, indicating either a grazing eclipse of a binary system or, more likely, a diluted background eclipsing binary.  In this case, the true orbital period is twice that identified by the TPS pipeline.   

The same type of astrophysical false-positive can be identified by searching for very shallow secondary events at phase=0.5. We must proceed with caution, however, in the interpretation of the secondary event since the occultation of a planet can also produce a secondary as is the case of HAT-P-7b \citep{hatp7b}.  If the depth of the secondary eclipse has a significance greater than 2-$\sigma$, then the dayside temperature of the planet candidate can be estimated.  The flux ratio of the planet and star ($F_p$/$F_*$) over the instrumental bandpass is given by the depth of the secondary eclipse.  The ratio of the planet and star radii is obtained from the transit depth. By assuming the star and companion behave as blackbodies and the flux ratio is bolometric, the dayside effective temperature can be estimated. 

For Kepler-5b (KOI-18; KeplerID 8191672) \citep{koi18} -- a confirmed planet with a 2.6-$\sigma$ secondary transit -- we find $F_p$/$F_*=3.3 \times 10 ^{-5}$, which gives $T_{\rm eff}=1657\pm223$ where an error of 30\% is assumed for the input stellar luminosity and radius. This estimate is a lower limit as a significant fraction of the planetary flux is emitted at wavelengths longer than the red edge of the {\em Kepler} bandpass, but it is a useful diagnostic to determine whether the depth of the secondary eclipse is consistent with a strongly irradiated planet. To make this comparison, we estimate the equilibrium temperature,
\begin{equation}\label{eq:teq}
T_{eq}=T_* (R_*/2a)^{1/2} [f(1-A_B)]^{1/4},
\end{equation}
for the companion, where $R_*$, $T_*$ are the stellar radius and temperature, with the planet at distance $a$ with a Bond albedo of $A_B$, and $f$ is a proxy for atmospheric thermal circulation. We assume $A_B = 0.1$ for highly irradiated planets \citep{row06} and $f=1$ for efficient heat distribution to the night side. These choices give a rough estimate for the dayside temperature of the planet assuming stellar irradiation is the primary energy source. Assuming a 30\% error in the input stellar parameters and that star and planet act as blackbodies we find $T_{eq}=1868\pm284$ K. The consistency of $T_{\rm eff}$ and $T_{eq}$ to within 1-$\sigma$ demonstrates that the secondary eclipse is consistent with a strongly irradiated planet. If $T_{\rm eff}$ were found to be much larger than $T_{eq}$, then the companion is likely self-luminous and thus a stellar binary.

\begin{figure}
\begin{center}
\includegraphics{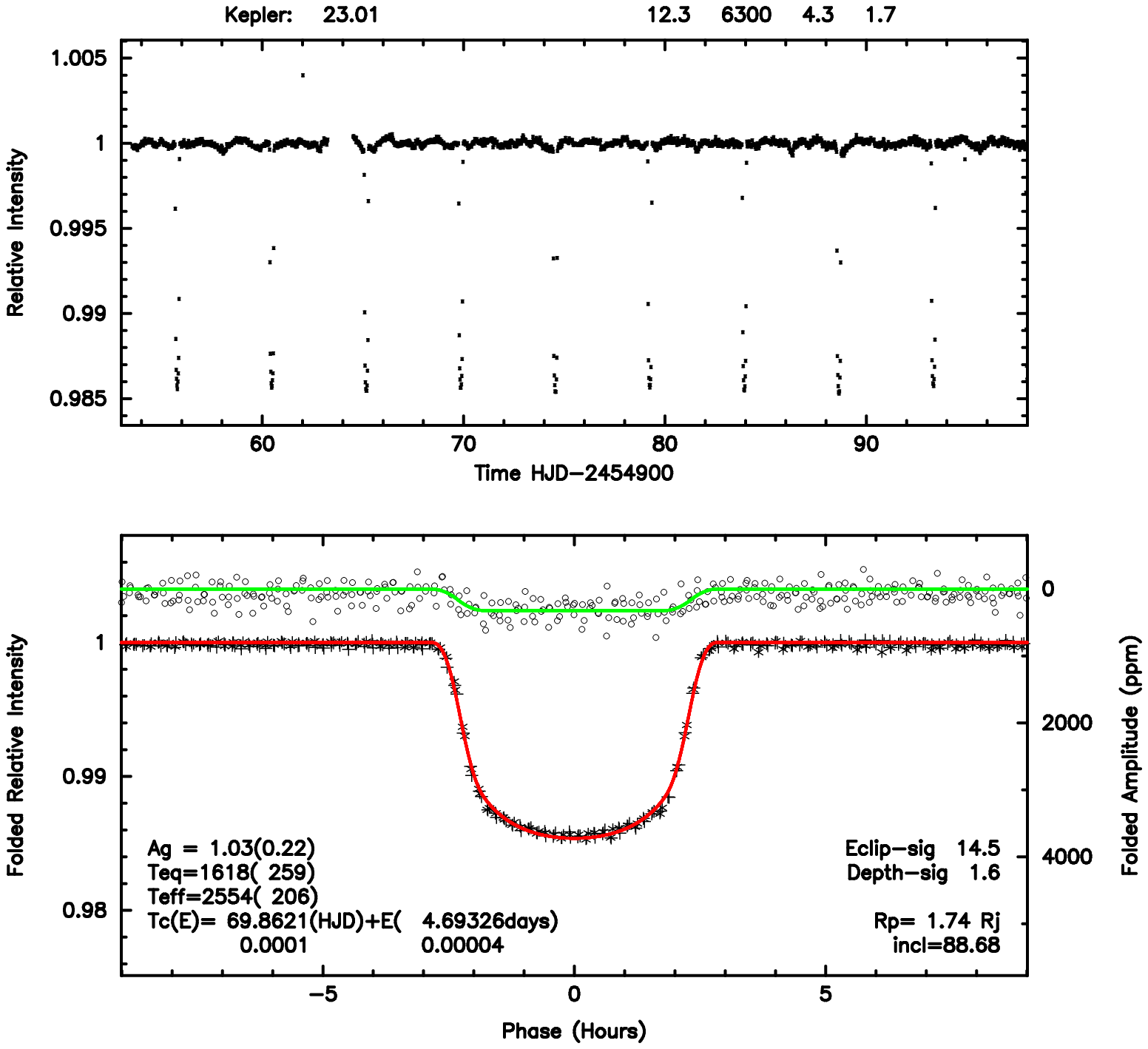}
\end{center}
\caption{Same as Figure~\ref{fig:oddEven} for KOI-23 (estimated companion radius, $R_p=1.74 R_J$). The 14-$\sigma$ (``Eclip-sig'') feature is not consistent with a planetary occultation as evidenced by the difference between the equilibrium temperature, $T_{eq}$, of a planet at that orbital period compared to the surface temperature, $T_{eff}$, required for a companion to yield the observed transit depth. One-sigma uncertainties for $T_{eq}$ and $T_{eff}$ are given in parentheses. These metrics are defined in Section~\ref{sec:modeling}. }
\label{fig:secEclipse}
\end{figure}

Figure~\ref{fig:secEclipse} illustrates the case of a secondary eclipse that is not consistent with occulted emission from a planet.  The light curve of KOI-23 (KeplerID 9071386), upper panel, as well as the phase folded transit, lower panel, is again shown.  The middle panel shows a close-up of the light curve at phase=0.5 where a 16-$\sigma$ secondary transit is evident.  Equation~\ref{eq:teq} yields $T_{eq}=1618\pm259$ K which is inconsistent with the effective temperature of the companion ($2554\pm206$ K) derived from the light curve (difference $>3.5$-$\sigma$).  The secondary transit of KOI-23 does not support the planetary companion interpretation.  This system is likely a grazing or diluted EB. We note, however, that the temperature argument is invalid if the planet is young or in an eccentric orbit.

\section{Photocenter Motion Diagnostics}
\label{sec:centroid}
Tracking the photocenter of the photometric aperture given {\em Kepler's} very high SNR and stable pointing is an effective means of identifying background eclipsing binaries.  The dimming of any object in the aperture will shift the photocenter of the light distribution since the photocenter is determined by the combination of various diffuse and discrete sources. The apparent change in the position of the target star due to a background eclipse event is dependent on the separation of the stars, their relative brightnesses, and the transit/eclipse depth.  A $50\%$ eclipse from a star 5 magnitudes dimmer than the target star and offset by one pixel will cause a 5 millipixel shift assuming 1) there are only two stars plus diffuse background in the photometric aperture and 2) all of the flux from the BGEB is included in the photometric aperture.  Pre-launch simulations of the photometric and astrometric stability of the instrument predict a 6.5-hour precision in the flux-weighted centroids for a 12th magnitude star of 20 $\mu$pix (80 $\mu$arcsec). 

To help determine whether background eclipsing binaries in the aperture of a KOI are responsible for the transit-like features identified by TPS, we correlate changes in centroid location with the transit-like features in the photometry. Systematics due primarily to focus and pointing changes \citep{pipeline,longcadence} were removed from the flux and centroid timeseries.  A high-pass filter was then applied to remove signatures occurring on timescales longer than 2 days.  Isolated outliers are removed from both the flux and centroid timeseries using a 5-sample wide moving median filter and rejecting points beyond 10 median absolute deviations from the moving median. 

\begin{figure}
\begin{center}
\includegraphics[height=7in,keepaspectratio=true]{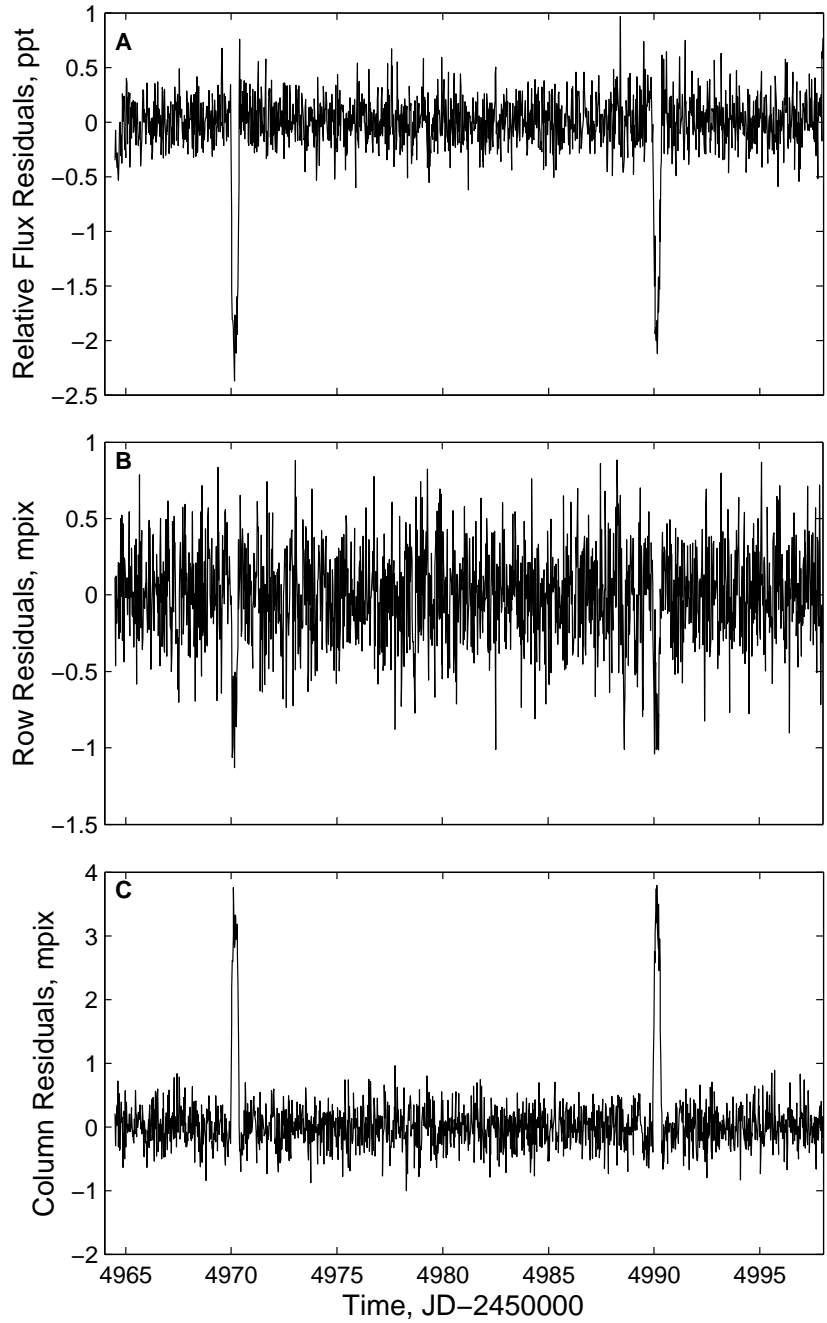}
\end{center}
\caption{Flux (upper, plotted in units of parts-per-thousand) and centroid (middle, lower) time series of KOI-140.  A high-pass filter has been applied to each timeseries to remove non-transit signatures on timescales longer than 2 days. Transit-like features in both the row and column centroid residuals are evident and highly correlated with the transit features in the flux time series suggesting this may be a diluted eclipsing binary.}
\label{fig:centrTimeSer}
\end{figure}

Figure~\ref{fig:centrTimeSer} shows the flux timeseries (upper) and the photocenter (flux-weighted centroid) timeseries of the row (middle) and column (lower) residuals for KOI-140 (KeplerID 5130369).  A 1 and 3 millipixel shift projected onto row and column, respectively, and correlated with the transit signal is clearly seen even at the 30-minute cadence.  An alternate way of displaying the same information is shown in Figure~\ref{fig:rainPlot}.  Here, the fractional change in brightness is plotted against the residual pixel position of the flux-weighted centroids, with row and column values distinguished using different symbols.  The ``cloud'' of points near (0,0) represents the out-of-transit points from which we estimate the per-cadence uncertainty -- 0.3 millipixels in the case of KOI-140.  The points that rain down are the in-transit points.  If the target star is responsible for the transits and if the photometric aperture is free from contaminating flux spatially offset from the target star, then the rain will fall vertically under the cloud corresponding to zero centroid residuals.  In the case of KOI-140, the rain falls diagonally, showing a $-0.68$ millipixel row residual and a $+3.02$ millipixel column residual. With 27 in-transit samples, this corresponds to a 12-$\sigma$ and 52-$\sigma$ significance, respectively.  These centroid offsets together with knowledge of the apparent brightnesses of stars in the vicinity (from the KIC) allow us to identify the star responsible for the transit-like event. 

In crowded fields, the flux distribution in a photometric aperture is rarely limited to the target star plus diffuse background.  Complex spatial flux distributions imply that the out-of-transit photocenter is not necessarily centered on the target star. Consequently, the photocenter can shift in unexpected directions during transit even when it is the target star itself that is responsible for the signal.  For this reason, we look towards other diagnostics in an effort to confirm the identity of the transiting (or eclipsing) object.

\begin{figure}
\begin{center}
\includegraphics[width=6in,keepaspectratio=true]{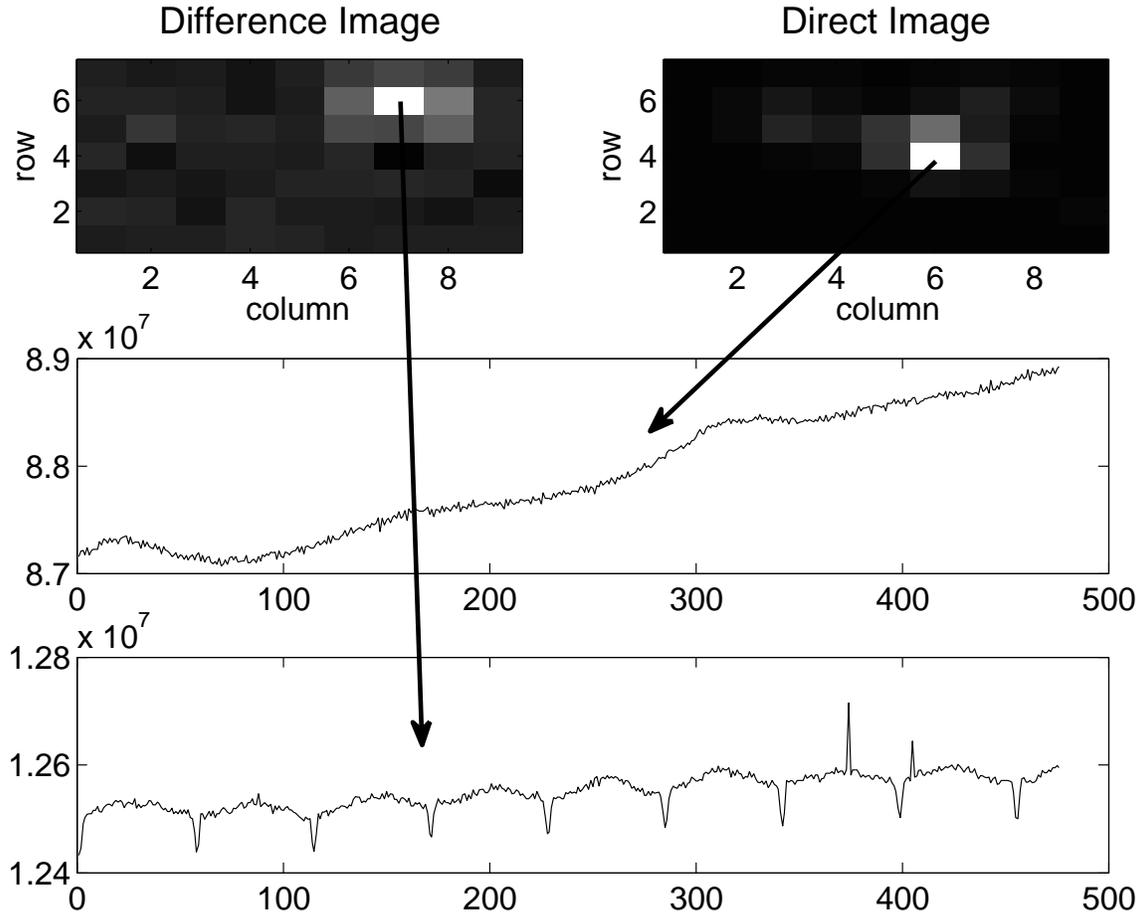}
\end{center}
\caption{An average of images taken at a time when KOI-08 is not transiting is subtracted from an average of images taken during transit.  The brightest pixel in the difference image (row=6, column=7) indicates the location of largest flux change and is not coincident with the position of the target star evident in the direct image (row=4, column=6).The corresponding pixel flux timeseries (both taken from a direct image during transit) confirms that the events detected in KOI-08 are due to a BGEB (middle and lower panel).}
\label{fig:diffim}
\end{figure}

KOI-08 (KeplerID 5903312) is an example of a BGEB.  The aperture is especially crowded, with 8 stars less than 3 magnitudes fainter than the target located within a 3-pixel radius.  The centroid timeseries and rain plot show a $\sim$0.75 millipixel centroid shift in both row and column.  An average of the images taken at cadences occurring during transit is subtracted from an out-of-transit image.  The upper panel of Figure~\ref{fig:diffim} shows the comparison of the out-of-transit image of KOI-08 (right) with the differenced image (left).  The brightest pixel in the difference image shows the spatial location of the pixels contributing most strongly to the flux change during the transit.  In this case, the position is offset by approximately one pixel in both row and column from the nominal target.  This coincides with an object in the KIC that is 1.5 magnitudes fainter in the {\em Kepler} passband than the target star (2.45 magnitudes in 2MASS J).  The lower panel shows two single-pixel flux timeseries.  The upper corresponds to the brightest pixel in the direct image, and the lower corresponds to the brightest pixel in the difference image.  The transit event disappears completely from the pixel timeseries drawn from the target star.  To the contrary, the fractional transit depth is larger in the flux timeseries of the brightest pixel in the difference image.    The centroid timeseries, the rain plot, the difference image, and the pixel flux timeseries collectively allow us to identify the majority of BGEBs directly from {\em Kepler} data without the need for more complex modeling or observations.

\section{Summary}
\label{sec:summary}

Modeling light curves under the assumption that the companion is a planet provides discriminators against grazing and diluted eclipsing binaries.  The metrics used to flag likely false-positives are the odd/even transit depth statistics and the secondary eclipse statistic.  The latter is complimented by a comparison of the equilibrium temperature of the planet computed from the orbial and stellar characteristics versus the day-side temperature of the planet computed from the transit/occultation depths.  When the two are markedly different, the planet interpretation for the secondary is discarded.  

\begin{figure}
\begin{center}
\includegraphics{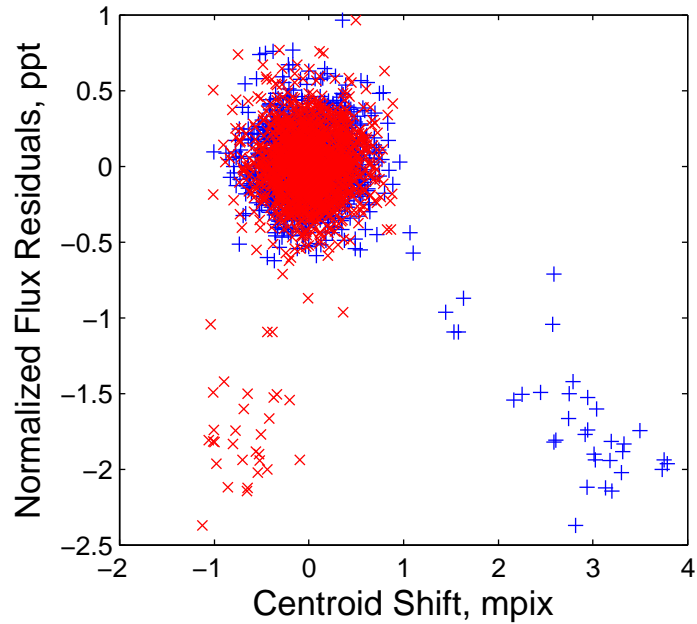}
\end{center}
\caption{Flux versus residual row (x) and column (+) centroids for KOI-140. The measurements are the same as those presented in the Figure~\ref{fig:centrTimeSer}. This ``rain plot'' shows that centroids are systematically shifting during each transit event.  The extent of the shift in each dimension, coupled to the depth of the transit-like feature is indicative of a BGEB. Had the transit originated from the central target star, the out-of-transit points would have rained down vertically under the cloud of out-of-transit points.} 
\label{fig:rainPlot}
\end{figure}

The centroid timeseries is a powerful discriminator against diluted eclipsing binaries.  KOI-140 is presented as an example of a BGEB identified via centroid analysis.  This 13.8-magnitude star yields a per-cadence (30-min) centroid precision of 0.3 millipixels, or 83 $\mu$pix at 6.5-hours. Rain plots exemplified in Figure~\ref{fig:rainPlot} provide an efficient visual assessment of the likelihood of a BGEB.  For more complex flux distributions in the photometric aperture, the difference image (out-of-transit minus in-transit) often clearly shows the location of the BGEB.  Individual pixel flux timeseries confirm the source in that they isolate the flux of the transiting system, minimizing dilution, and thereby diminishing (or augmenting) the transit depth. This is effective when the BGEB is spatially well-separated from the target star and the stars are sufficiently bright.
 
All {\em Kepler Objects of Interest} must pass the modeling and centroid inspections before being passed on for follow-up observations.  In the upcoming year, these metrics will be folded into pipeline Data Validation tools forming a more efficient TCE-to-KOI filter.

\acknowledgments

The authors wish to acknowledge the {\em Kepler Science Operations Center}, especially Hema Chandrasekharan for her work on the transiting planet search software.  Funding for this Discovery mission is provided by NASA's Science Mission Directorate.


\clearpage 

\clearpage

\clearpage

\clearpage

\clearpage

\end{document}